\documentclass[preprint2]{aastex}

\usepackage{ulem}
\def\fb#1  {{\bf [FB:~ #1]~}}
\def\js#1  {{\bf [JS:~ #1]~}}
\def\new#1 {{\bf #1 }}
\def\cut#1 {\sout{#1} }

\def\spose#1{\hbox to 0pt{#1\hss}}
\def\simlt{\mathrel{\spose{\lower 3pt\hbox{$\mathchar"218$}}
     \raise 2.0pt\hbox{$\mathchar"13C$}}}
\def\simgt{\mathrel{\spose{\lower 3pt\hbox{$\mathchar"218$}}
     \raise 2.0pt\hbox{$\mathchar"13E$}}}

\slugcomment{Rev. 1 - 2005 April 28}
\shorttitle{Near-IR Photometry of RD0301}
\shortauthors{Staguhn et al.}

\begin{document}

\title{Near-Infrared Photometry of the High-Redshift Quasar
RDJ030117+002025: Evidence for a Massive Starburst at $z=5.5$\altaffilmark{1}}

\author{J.G. Staguhn\altaffilmark{2,3},  D. Stern\altaffilmark{4}, D.J.
Benford\altaffilmark{2}, F. Bertoldi\altaffilmark{5}, S.G.
Djorgovski\altaffilmark{6}, D. Thompson\altaffilmark{6}}

\email{staguhn@stars.gsfc.nasa.gov}
\altaffiltext{1}{Based on the observations obtained at the W.M. Keck Observatory, which is
operated as a scientific partnership among the California Institute of
Technology, the University of California and the National Aeronautics and Space
Administration. The Observatory was made possible by the generous financial 
support of the W.M. Keck Foundation.}
\altaffiltext{2}{Observational Cosmology Laboratory (Code 665), NASA/Goddard Space Flight Center, Greenbelt, MD 20771, USA}
\altaffiltext{3}{SSAI, Lanham, MD 20706, USA}
\altaffiltext{4}{JPL/Caltech, MS 169-506, Pasadena, CA 91109, USA}
\altaffiltext{5}{Radioastronomisches Institut, Universit\"at Bonn, Auf dem H\"ugel 71, 53121 Bonn, Germany}
\altaffiltext{6}{Division of Physics, Mathematics, and Astronomy, Caltech, Pasadena, CA 91125, USA}

\begin{abstract}

With a redshift of $z=5.5$ and an optical blue magnitude $M_B \sim -24.2$ mag ($\sim 4.5
 \times 10^{12}$~$L_\sun$), RDJ030117+002025 is  the most distant optically
 faint ($M_B > -26$ mag) quasar known.  MAMBO continuum observations at
 $\lambda=1.2$ mm ($185 \mu$m rest-frame) showed that this quasar has a far-IR
 luminosity comparable to its optical luminosity. 
We present near-infrared $J$- and
 $K$-band photometry obtained with NIRC on the Keck~I telescope, tracing the
 slope of the rest frame UV spectrum of this quasar.  The
 observed spectral index is close to the value of  $\alpha_\nu \sim -0.44$
 measured in composite spectra of optically-bright SDSS quasars. It thus appears
 that the quasar
 does not suffer from strong dust extinction, which further implies that its low
 rest-frame UV luminosity is due to an intrinsically-faint AGN. The FIR to
 optical luminosity ratio is then much larger than that observed for the
 more luminous quasars, supporting the suggestion that the FIR emission is not
 powered by the AGN but by a massive starburst.
\end{abstract}


\section{Introduction}

Recent {\it Hubble Space Telescope} studies have shown that most
spheroidal galaxies in the nearby universe contain supermassive black holes (SMBHs),
and that there is a strong correlation between the black hole mass
and the velocity dispersion of the stars \citep{geb00}, as well as other fundamental properties of galaxies;
 see, e.g., \citet{fer02} for a review.
At the same time, there are good reasons to believe that the same
phenomenon -- dissipative merging and interactions -- can power both
starbursts and the central activity in galaxies, leading to a joint
formation and co-evolution of galaxies and SMBHs; see, e.g., \citet{hae04}, or \citet{djo04}, for reviews and references.

The identification of high-$z$ quasars is still largely
the domain of optical large-area sky surveys, 
but because of dust extinction it is
prone to selection effects. All but two of the quasars known to date
at $z>5$ were identified in optical surveys with observed blue rest-frame magnitude
detection limits  $M_B \simlt -26$ mag. These quasars are
 optically extremely luminous, and presumably have unusually
small foreground dust and associated extinction. 
Our knowledge about the nature and the properties of individual
high-$z$ quasars mostly comes from follow-up observations
of these bright, optically-selected sources at X-ray, radio, mm- and
submm wavelengths.  Millimeter surveys of optically bright-quasars at
high redshifts ($z \simgt 2$) show that about one-third of 
all bright quasars
have FIR luminosities of order $10^{13}L_\sun$, making them detectable to
  current (sub)mm bolometer detectors 
 \citep{omo01, car01, omo03}. 
The thermal
nature for the observed FIR emission was implied by 
the  observer-frame
cm-to-mm spectral indices and the fact that high molecular gas masses
are detected in a number of high-$z$ quasars \citep[][and references
therein]{wal03}. 

It remains unclear to what extent  AGN do contribute to the FIR luminosity of quasars. Spectral information on the
mid-IR SED remains scarce for high-z quasars.

%
A way to support the mounting evidence for a starburst origin of the FIR emission
 is to establish the lack of a correlation between the optical
 to mid-IR luminosity (which clearly derives from the AGN) and the FIR
 luminosity.
In a sample of SDSS and PSS quasars, \citet{omo01} and \citet{ber02} find no
clear correlation between observer frame mm- and optical luminosity.
A similar conclusion is drawn by \citet{pri03} who observed a sample of 57 optically selected
quasars with $M_B < -27.5$ mag  and at $z \sim 2$ in the submm with SCUBA.
%
%
%
%

\subsection{RDJ030117+002025} \label{RD0301_Intro}


At $z=5.5$ and with $M_B \sim -24.2$ mag (using the WMAP cosmological parameters from \citet{spe03}),  RD~J030117+002025 (hereafter RD0301) is  the most distant quasar identified to date with 
$M_B > -26$ mag. It was discovered by \citet{ste00} as a ``$R$-band dropout"  from deep $RIz$-band imaging in a 74 arcmin$^2$  field, observed for the SPICES survey \citep{ste01}.  At $z\simeq5$ high redshift galaxies and quasars disappear from the $R$-band, because the
Ly$\alpha$ and Ly$\beta$ forests attenuate the rest-frame ultraviolet continuum.
The optical luminosity of RD0301 is  $\sim 4.5\times10^{12} L_\sun$ (we apply the same method for the conversion of absolute blue magnitude into optical luminosity as described in \citet{ber02}).  The spectrum of RD0301 shows broad Ly$\alpha$/Nv
$\lambda 1240$  emission \citep{ste00}. 
Compared to the typical SDSS- or PSS-selected source, RD0301 is 
approximately four magnitudes fainter at optical wavelengths.
A MAMBO 1.2~mm continuum detection ($185 \mu$m rest-frame) 
shows that the host galaxy is an ultraluminous
FIR source \citep{ber02}: with a 1.2~mm continuum flux density of
$0.87\pm0.20$~mJy, its FIR luminosity is $L_{\rm{FIR}} \sim 5\times10^{12}
L_\sun$ , which is comparable to the average FIR luminosity of 
the optically-bright, high-$z$ quasars observed at mm and submm wavelegths. 
The observed VLA A-array 1.4 GHz flux density of $73 \pm 17
\mu$Jy \citep{pet03} coincides positionally within 0\farcs5 with the optical position of RD0301,
and is consistent  with a thermal nature of the
FIR rest-frame luminosity \citep{con92}. The high angular resolution of the 1.4 GHz detection, combined with the sensitivity of these observations, allows us to exclude that the observed 1.2~mm emission is from a spatially distinct dusty companion galaxy, which might be undetected in the optical observations. 

\section{Observations and Data Reduction}
\label{nircobs}


Infrared images of RD0301 were obtained in the $J$ and $K$ band filters
on UT 2004 August 06 with the Near-Infrared Camera \citep{nirc94} on
the Keck~I telescope.   NIRC reimages a 38\farcs4 square field
of view onto a 256$^{2}$ InSb detector at 0\farcs15 pixel$^{-1}$.
The observations were obtained under photometric conditions, and the
seeing was 0\farcs36 FWHM in $K$ and 0\farcs47 in $J$.

The data consist of multiple spatially-dithered images in each filter,
with exposure times chosen as appropriate for the background flux and
coadds chosen to bring the total exposure per pointing to one minute.
Using IRAF\footnote{IRAF is distributed by the National Optical Astronomy
Observatories, which are operated by the Association of Universities for
Research in Astronomy, Inc., under cooperative agreement with the National
Science Foundation.} routines, the individual images were dark subtracted,
flatfielded, masked, aligned, and stacked with integer pixel shifts.
Several $J$-band images were not used for lack of sources on which to
centroid, resulting in final exposure times of 10 minutes in $J$ and 12
minutes in $K$.

The data were calibrated onto the Vega scale using the \citet{pers98}
standard stars P530-D and P533-D, observed immediately before and after
the quasar and at similar airmass.  Photometry on the standards was
extracted in a 4\arcsec\, diameter aperture.  The quasar photometry was
extracted in a 2\arcsec\, diameter aperture and corrected to 4\arcsec\,
using aperture corrections derived from a brighter star in the field.  This relatively small aperture
was used to maximize the signal-to-noise ratio on the quasar flux.  The
photometric uncertainties are dominated by the source flux measurement,
but include the zero point uncertainties from the standard stars.
We obtain magnitudes of $J_{\rm AB} = 22.98 \pm 0.18$ and $K_{\rm AB} =
22.6 \pm 0.19$ for the quasar.  The corresponding Vega-based magnitudes
are $J_{\rm Vega} = 22.08 \pm 0.18$ and $K_{\rm Vega} = 20.7 \pm 0.19$.

\section{Results and Discussion}

The  goal of our near-IR observations was to determine whether
the  (rest-frame) UV faintness  of RD0301 is due to the 
foreground extinction of a bright AGN, or whether the AGN is
intrinsically faint.  


By expanding the observed spectral energy distribution into the near-IR
we are able to better determine the rest-frame UV spectral index,
and thereby place constraints on the extinction toward the AGN.
Recently, \citet{mai04} found evidence for a
different extinction curve blueward of 1700 \AA ~ toward a $z = 6.2$ BAL quasar.
This is the first evidence for a change of  dust properties toward high redshift objects.  We therefore aimed at a
measurement redward of 1700 \AA\ to avoid possible ambiguities in the interpretation of the observed extinction toward  the AGN.

The observed ${J-K}$ color index of $0.38 \pm 0.26$ is, within the
uncertainty, identical to the color obtained by redshifting the LBQ
composite quasar spectrum \citep{fra91} to $z = 5.50$, which yields ${J-K}$(LBQ)
$= 0.28$.  It is also identical to the color derived from the redshifted
SDSS composite spectrum \citep{van01}, shown in Figure~1. The template
spectrum shown here was constructed from
over 2200 SDSS spectra. Figure~1 demonstrates that the UV rest-frame continuum emission
of RD0301 is  consistent with  that of the
standard quasar spectrum, which has a spectral index of $\alpha_\nu \sim
-0.44$.  In particular the $r$, $z$, $J$, and $K$-band data points coincide
with the expected values.
Only the measured $I$-band luminosity deviates significantly from the expected value;
however, this band coincides with the Ly$\alpha$ emission line and
therefore is expected to show the largest variations.
We also note that the RD0301 photometry shortward of $1\mu$m 
was obtained 
 several years before the near-IR photometry presented here; any quasar variability could be a source of
additional uncertainty in our interpretation of the quasar SED. 

Significant obscuration, as is present in most optically-faint, X-ray
selected sources,  would lead to notably steeper spectral indices. Typical spectral indices longward of Ly$\alpha$ of such sources range from $\alpha_\nu = -0.54$ to $\alpha_\nu = -2.69$ \citep{wil04}.  Using the uncertainties in the $J$-band photometry, and applying the dust extinction laws for the Milky Way and the Small Magellanic Cloud (SMC) found by \citet{pei92}, we obtain an upper limit for the color excess  with respect to the template spectrum of unobscured quasars of E(B-V) $\simlt 0.04$ (Milky Way) and E(B-V) $\simlt 0.03$ (SMC). This corresponds to an upper limit of $2\times 10^{20} N_{\rm{H{\sc I}}}$ cm$^{-2}$  (Milky Way), and   $1\times 10^{21} N_{\rm{H{\sc I}}}$ cm$^{-2}$  (SMC) of foreground neutral hydrogen columns over the unobscured template quasar. The dust color excess to neutral hydrogen relationship for the SMC is probably more applicable for high-$z$ quasars \citep{wil04}.  The prior derivation would not be valid under either of the following conditions: a) the foreground dust is grey over the wavelength range of the photometry presented here (for a review of our current understanding of dust models, see \citet{dwe05}) or b) the foreground dust is optically thick and the observed UV emission from the quasar is reflected from dust clouds. If both of those conditions are not present we  can exclude that foreground extinction reduces the observed blue magnitude of RD0301 by the 4 magnitudes that it is fainter than the average luminous SDSS high-$z$ quasars. The inclination of the host galaxy and the geometrical  distribution of the dust in the host galaxy  appears to allow an unobscured view of the quasar, as is the case with the more luminous high-z quasars (``Type I" quasars). The AGN then must be intrinsically fainter by $\sim 4$ mag than SDSS and PSS high-$z$ quasars.

With $L_{\rm FIR} \simeq L_{\rm B}$ and  the AGN being
unobscured, it is still possible that the FIR luminosity
of RD0301 is mainly powered by the AGN. However, if this were the case
one would need to explain why optically-bright high-$z$ quasars
typically have  $L_{\rm FIR} \simlt 0.1 \times L_{\rm B}$ \citep[Figure~2
in][]{ber02}, since this would imply a significantly lower dust heating
efficiency of the AGN in optically-bright quasars.  Furthermore, even for these optically bright AGN there is no clear correlation between optical and FIR luminosity, suggesting there is no causal relation. Certain geometries (and quantities) of the dust around
the AGN, that would have to be distinct from all high-$z$ SDSS and PSS quasars,  might be able to account for the observed differences.  However, the fact that R0301 appears unobscured strengthens the
suggestion that the dust is not in the immediate surrounding and
the FIR emission is therefore not powered by the AGN. A massive starburst
as the major heating source of the FIR emission of RD0301 seems more likely.



\begin{figure}
\plotone{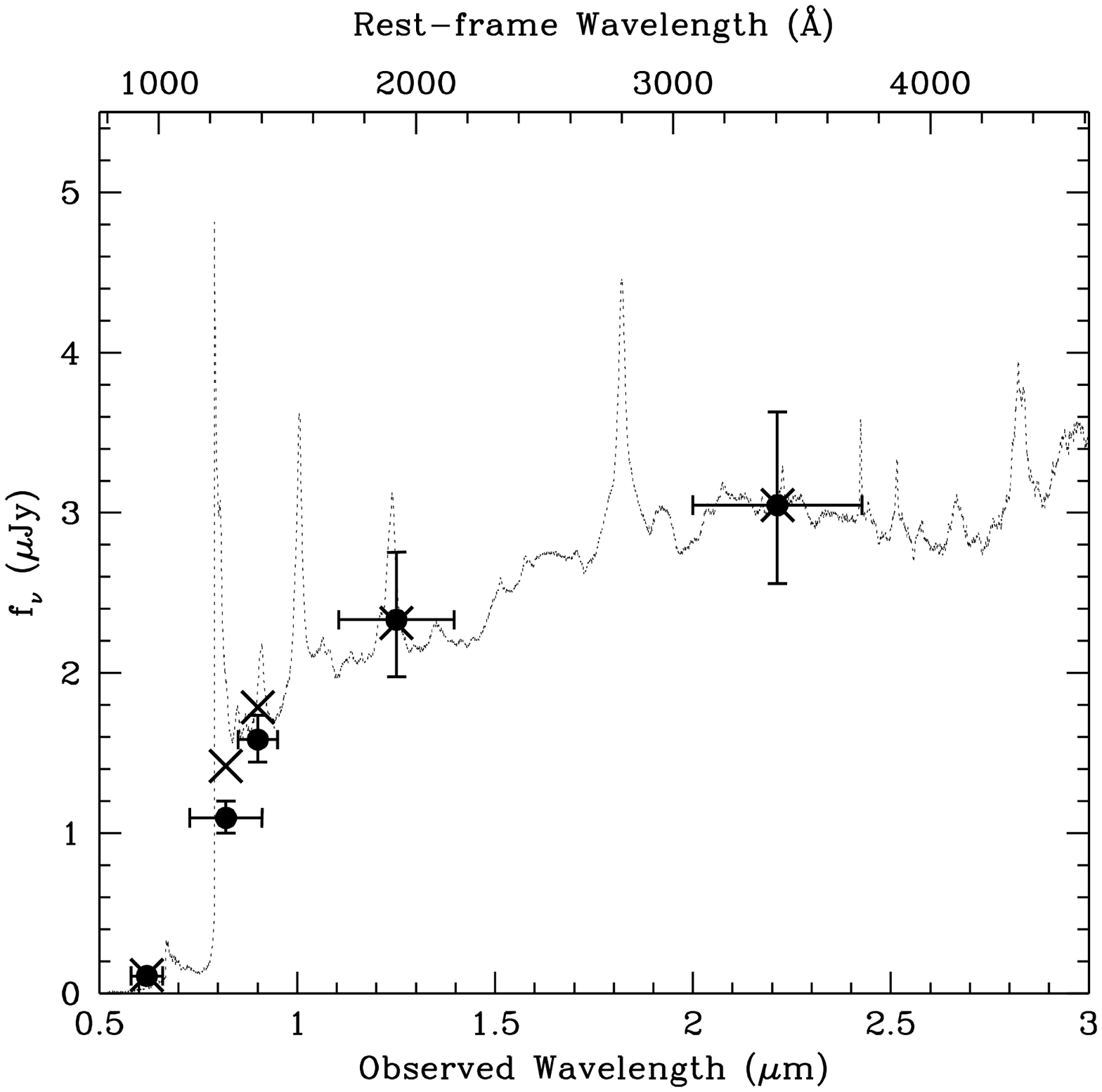}
\caption{Spectral energy distribution of RD0301.  Filled circles with error bars show the
photometry of the source.  Points redward of 1~$\mu$m are from this paper;
points blueward of 1~$\mu$m are from \citet{ste00}.  The dotted line shows
the \citet{van01} composite quasar spectrum, constructed
from over 2200 Sloan Digital Sky Survey spectra, shifted to redshift
$z = 5.50$ and normalized to match the observed $K$-band flux density.
Absorption according to the \citet{mad95} parameterization of the Lyman
forests has been applied.  Crosses indicate the predicted brightness of
the composite quasar at the observed filters.  RD0301 appears to be a
normal quasar in the rest-frame ultraviolet, virtually indistinguishable
from the composite quasar spectrum.}
\label{fig1}
\end{figure}

\section{Conclusion and Outlook}

The near-IR photometry of RD0301 presented here
implies that  the observed relatively faint absolute blue magnitude M$_B \sim  -24.2$ mag  of this $z=5.5$ quasar 
is intrinsic to the AGN, which translates into the quasar having about the same intrinsic optical luminosity as its restframe FIR luminosity of $ \sim 5\times10^{12} L_\sun$. 
If the AGN were the dominant heating 
source of the observed FIR luminosity, then its dust heating efficiency would be 
required to be a factor of at least 10 higher than the upper limit found in SDSS and
PSS high-z quasars \citep{ber02}. The most likely scenario therefore is that the observed FIR luminosity  originates from thermal dust emission heated by a massive starburst in the host galaxy of this quasar.  Such massive starbursts could account for the richness of metals \citep{omo96,ber03} and the large gas reservoirs \citep{oht96,ham99,wal03,die03,cox05} found in the earliest
quasars known to date. 

X-ray observations, which -- due to the high redshift -- could be
obtained with {\it Chandra} or {\it XMM} in the very hard rest-frame
bands between $\sim$ 2 and 52 keV, could be used to confirm the finding presented here.
Due to the lack of very deep, large-area optical surveys with sensitivities exceeding the SDSS and PSS surveys by several orders of magnitude we are currently not able to establish the space density of 
faint, high-$z$ quasars with similar properties in the early universe. 

\acknowledgements

The authors wish to recognize and acknowledge the very significant
cultural role and reverence that the summit of Mauna Kea has always
had within the indigenous Hawaiian community; we are most fortunate
to have the opportunity to conduct observations from this mountain.
The work of DS was carried out at Jet Propulsion Laboratory, California
Institute of Technology, under a contract with NASA.

{}
\end{document}